\documentclass[english]{iopart_mod}

\usepackage[reqno,centertags,intlimits,fleqn]{amsmath}
\usepackage{amsfonts}
\usepackage{amssymb}
\usepackage{babel}
\usepackage{cite}
\usepackage{graphicx}

\usepackage{ASRHD}
\begin{document}

\paper{Asymptotic structure of radiation in higher dimensions}

\author{Pavel Krtou\v{s} and Ji\v{r}\'{\i} Podolsk\'y}

\address{%%
  Institute of Theoretical Physics,
  Faculty of Mathematics and Physics,\\
  Charles University in Prague,\\
  V Hole\v{s}ovi\v{c}k\'{a}ch 2, 180 00 Prague 8, Czech Republic
  }
\eads{\mailto{Pavel.Krtous@mff.cuni.cz}, \mailto{Jiri.Podolsky@mff.cuni.cz}}

\date{December 6, 2005} %  accepted version
% version 2.02 (2006-02-02) - after 1st proofs
% version 2.03 (2006-02-13) - 2 references added - only for gr-qc

\begin{abstract}
We characterize a general gravitational field near conformal infinity (null, spacelike, or timelike)
in spacetimes of any dimension. This is based on an explicit evaluation of the dependence
of the radiative component of the Weyl tensor on the null direction
from which infinity is approached.  The behaviour similar to peeling
property is recovered, and it is shown that the directional structure
of radiation has a universal character that  is determined by
the  algebraic type of the spacetime. This is a natural generalization
of analogous results obtained previously in the four-dimensional case.
\end{abstract}

\submitto{\CQG}
% 04.20.Ha Asymptotic structure
% 04.50.+h Gravity in more than four dimensions, Kaluza-Klein theory, unified field theories; alternative theories of gravity
% 98.80.Jk Mathematical and relativistic aspects of cosmology

\maketitle

\section{Introduction}
\label{sc:introduction}

There has been a growing interest in studies of higher dimensional spacetimes,
mainly motivated by finding particular models in the contexts of string theory
and brane cosmology. However, some fundamental questions such as the mathematical
classification of manifolds based on the algebraic structure of the Riemann
and Weyl tensor, or investigation of the asymptotic behaviour of fields in
higher dimensions have only recently been initiated
\cite{ColeyMilsonPravdaPravdova:2004,PravdaPravdovaColeyMilson:2004,%
MilsonColeyPravdaPravdova:2005,HollandsWald:2004,HollandsIshibashi:2005,%
PravdovaPravdaColey:2005}, cf. also
\cite{DeSmet:2002,FrolovStojkovic:2003}.

As a contribution to this topic, in the present work we study asymptotic
properties of a gravitational field as represented by the Weyl tensor in
an arbitrary dimension. In particular, we analyze the directional structure at
conformal infinity of the leading component of the field which corresponds to radiation.
In fact, this is a natural extension of our previous work
\cite{KrtousPodolskyBicak:2003,KrtousPodolsky:2004a,KrtousPodolsky:2004b,KrtousPodolsky:2005a}
in which we completely described the asymptotic directional structure of radiation
in four-dimensional spacetimes with conformal infinity of any character (null, spacelike, or timelike).
We demonstrated that this directional structure has universal properties
that are basically given by the algebraic type of given spacetime,
namely the degeneracy and orientation of principal null directions of the Weyl tensor.
In the present article we show that these results---which are valid in
standard ${n=4}$ general relativity---can be directly generalized to higher dimensional spacetimes.
Below we prove that the asymptotic directional structure of gravitational
radiation in any dimension is given by the specific properties of Weyl aligned null directions
at conformal infinity, i.e., by algebraic type of spacetime at infinity.

The paper is organized as follows. In section~2 we introduce necessary
geometrical concepts and objects, and we set up the notation.
In section~3 we first summarize the algebraic classification of the Weyl tensor
in higher dimensions, and then we derive the expression which explicitly describes
the behaviour of the field at conformal infinity. Subsequently, we discuss the directional
structure of radiation in case of null, spacelike, and timelike infinity, in particular
for the simplest algebraically special spacetimes. In the appendix, the relation between
the higher-dimensional formalism used and the standard NP formalism in ${n=4}$ is presented.

For brevity, we refer to equations of the review paper \cite{KrtousPodolsky:2004b} directly
as, e.g., (R2.13).

\section{Geometrical preliminaries}
\label{sc:geom}

\subsection{Conformal infinity and null geodesics}
\label{ssc:confinf}
First, we briefly review the context in which we study the asymptotic behaviour of the gravitational field.
For details see the introductory section 2 of \cite{KrtousPodolsky:2004b}, where the discussion
was not restricted to a particular number of dimensions.

We wish to study spacetimes with a conformal infinity. We therefore assume
existence of an extension of the spacetime to an auxiliary manifold with metric ${\cmtrc}$
to which the physical metric ${\mtrc}$ is (at least locally) conformaly related by
${\cmtrc=\om^2\mtrc}$.
In the physical spacetime, the conformal factor ${\om}$ is positive; the hypersurface ${\om=0}$
corresponds to the spacetime infinity---called \defterm{conformal infinity  ${\scri}$}.
We assume regularity of the conformal geometry across ${\scri}$, even though it is known
that in higher dimensions this is a more subtle issue  than in the case ${n=4}$, cf.~\cite{HollandsWald:2004,HollandsIshibashi:2005}.
We will return to this question shortly in section~\ref{ssc:fieldasympt}.

We introduce a vector ${\norm}$ normal to ${\scri}$, ${\norm\propto\grad\om}$,
normalized using the physical metric,\footnote{%
The dot \vague{$\spr$} denotes a scalar product defined by the physical metric ${\mtrc}$.
Strictly speaking, at infinity we should define a normal ${\cnorm}$ normalized using conformal
geometry to which the vector ${\norm}$ is related by rescaling by ${\om}$ (which degenerates on ${\scri}$).
However, it is common  to use formally the normal ${\norm}$---see discussion in \cite{KrtousPodolsky:2004b}.}
${\norm\spr\norm=\nsgn}$,
where the constant factor ${\nsgn}$ indicates the character of infinity:
${\nsgn=-1}$ for spacelike ${\scri}$, ${\nsgn=0}$ for null ${\scri}$, and ${\nsgn=+1}$ for timelike ${\scri}$.
%\begin{equation}
%\nsgn=
%\begin{cases}
%   -1: &       \quad \ text{$\scri\,$ is \emph{spacelike},}  \cr
%   \hfill 0: & \quad \text{$\scri\,$ is \emph{null},} \cr
%   +1: &       \quad \text{$\scri\,$ is \emph{timelike}.} \cr
%\end{cases}
%\label{DefinitionSigma}\end{equation}
From Einstein's field equations, assuming a vanishing trace of the energy-momentum tensor,
it follows \cite{PenroseRindler:book,KrtousPodolsky:2004b}
that the character of infinity is correlated with the sign of cosmological
constant, ${\nsgn=-\sign{\Lambda}}$.

By radiative component we understand
the leading component of the field measured with respect of
a specific frame along a future oriented \emph{null} geodesic ${\geod(\afp})$ approaching
${\scri}$. We are interested
in the dependence of such a component on a \emph{direction}
along which infinity is approached. To compare the field
along different geodesics, we have to fix the normalization of the affine
parameter of these geodesics. We require that the projection of the tangent vector
of the geodesic to ${\norm}$ is independent of the direction
of the geodesic. Using the relation to conformal geometry it can be
shown (see (R2.13), (R2.14)) that near the infinity we have
\begin{equation}\label{InOutGeod}
%\frac{D\geod^a}{d\,\afp}\,\grad_a\om\bigg|_\scri\equiv
\frac{d\,\om}{d\,\afp}\lteq-\EPS\,\om^2 \period
\end{equation}
Here, the sign ${\EPS}$ characterizes the orientation of the geodesic with respect
to conformal infinity:
\begin{equation}
\EPS=
\begin{cases}
   +1: &       \quad \text{for \emph{\,outgoing geodesics},}\quad \afp\to+\infty \text{\, on } \scri, \cr
   -1: &       \quad \text{for \emph{\,ingoing geodesics},\,}\ \quad \afp\to-\infty \text{\, on } \scri. \cr
\end{cases}
\label{DefinitionEpsilon}\end{equation}
From \eqref{InOutGeod} it follows that
\begin{equation}\label{OmegaInAfp}
\om=\EPS\,\afp^{-1}+\dots\period
\end{equation}

\subsection{Null frames and their transformations}
\label{ssc:nullframes}
In four dimensions, it is convenient to introduce complex null tetrads (R3.1).
In the case of higher dimensions, we have to choose a slightly different normalization of the vectors of a real frame.
Following \cite{MilsonColeyPravdaPravdova:2005}, we call the frame ${\kG,\,\lG,\,\mG{i}}$
the \defterm{null frame} if ${\kG}$, ${\lG}$ are future oriented null vectors, and ${\mG{i}}$, ${i=1,2,\dots,n-2}$,
are spatial real vectors satisfying
\begin{equation}\label{NullFrNorm}
   \kG\spr\lG = -1\comma \kG\spr\mG{i}=0\comma \lG\spr\mG{i}=0\comma
   \mG{i}\spr\mG{j} = \krdel_{ij} \period
\end{equation}
We use indices ${a,b,c,\dots}$ to refer to all spacetime dimensions,
and indices ${{i,j,k,\dots}=1,2,\dots,n-2}$ to label spatial directions orthogonal to ${\kG,\lG}$.
Thanks to the orthonormality relation \eqref{NullFrNorm},
components of any spatial vector ${\tens{V}}$ spanned on the vectors
${\mG{i}}$, i.e. ${\tens{V}=V^i\mG{i}}$, satisfy ${V^i=V_i}$.
We also use a standard shorthand for square of the magnitude ${\abs{V}^2=\tens{V}\spr\tens{V}=V^i V_i}$.

We denote the vectors of an associated \defterm{orthonormal frame} as
${\tG,\,\qG,\,\mG{i}}$, where
\begin{equation}\label{OrthFr}
  \tG = \textstyle{\frac1{\sqrt{2}}} (\kG+\lG)\comma
  \qG = \textstyle{\frac1{\sqrt{2}}} (\kG-\lG)\period
\end{equation}

We will distinguish different null frames by an additional lower roman index.
For example, below in section \ref{ssc:RefFrame} we will introduce
\defterm{reference frame} denoted as ${\kO,\,\lO,\,\mO{i}}$.
General transformations between different null frames can be composed from
the following simple Lorentz transformations:
\begin{itemize}
\item\defterm{null rotation with ${\kG}$ fixed} (parametrized by a spatial vector ${\Lv=L^i\mG{i}}$):\footnote%
{Note that our parametrization of null rotations differs from that used in \cite{MilsonColeyPravdaPravdova:2005} by factor ${\sqrt2}$.}
\begin{equation}\label{kfixed}
  \kG = \kO\commae\qquad
  \lG = \lO + \sqrt2 L^i \mO{i} + \abs{L}^2\, \kO\commae\qquad
  \mG{i} = \mO{i} + \sqrt2\, L_i\,\kO\commae
\end{equation}
\item\defterm{null rotation with ${\lG}$ fixed} (parametrized by a vector ${\Kv=K^i\mG{i}}$):
\begin{equation}\label{lfixed}
  \kG = \kO + \sqrt2 K^i \mO{i} + \abs{K}^2\, \lO\commae\qquad
  \lG = \lO \commae\qquad
  \mG{i} = \mO{i} + \sqrt2\, K_i\,\lO\commae
\end{equation}
\item\defterm{boost in the ${\kG\textdash\lG}$ plane} (parametrized by a real number ${B}$):
\begin{equation}\label{boost}
  \kG = B\,\kO \comma\quad
  \lG = B^{-1}\, \lO \comma\quad
  \mG{i} = \mO{i} \commae
\end{equation}
\item\defterm{spatial rotation in the space spaned on ${\mG{i}}$} (parametrized by an orthogonal matrix ${\Phi_i{}^j}$):
\begin{equation}\label{rotation}
  \kG = \kO \comma\quad
  \lG = \lO \comma\quad
  \mG{i} = \Phi_i{}^j\,\mO{j} \comma\quad\text{with}\quad
  \Phi_i{}^j\,\Phi_k{}^l\;\krdel_{jl}=\krdel_{ik}\period
\end{equation}
\end{itemize}

We say that a null frame is \defterm{adjusted to conformal infinity ${\scri}$}\
if the null vectors $\kG$ and $\lG$  on $\scri\,$ are coplanar with normal $\norm$
to the conformal infinity, and they satisfy the relation
\begin{equation}\label{Adjusted}
   \norm = \EPS\textstyle{\frac{1}{\sqrt{2}}}
   (-\nsgn\kG+\lG)\comma\quad\text{where}\quad \EPS=\pm1\period
\end{equation}
It follows that for a spacelike infinity (${\nsgn=-1}$)
${\norm = \EPS\,\tG}$, for a timelike $\scri\,$ (${\nsgn=+1}$)
 ${\norm = -\EPS\,\qG}$, and ${\norm = \EPS\,\lG/\sqrt{2}\,}$
for null $\scri\,$ (${\nsgn=0}$).
Clearly, the vectors ${\mG{i}}$ of the adjusted frame are tangent to ${\scri}$.
If the null vector $\kG$ is oriented along the null geodesic
$\geod(\afp)$, the parameter $\EPS$ indicates whether the geodesic is outgoing (${\EPS=+1}$)
or ingoing (${\EPS=-1}$) (cf.\ fig.~2 of \cite{KrtousPodolsky:2004b}).

\subsection{The reference frame and parametrization of null directions}
\label{ssc:RefFrame}

To parametrize a null direction along which ${\scri}$ is approached, we
fix at conformal infinity  a reference frame ${\kO,\,\lO,\,\mO{i}}$.
We require that this is adjusted to infinity in the sense of \eqref{Adjusted}
and that it is smooth\footnote{Again, at infinity we should define the frame
normalized in conformal geometry to which the frame ${\kO,\,\lO,\,\mO{i}}$
is related by isotropic rescaling by ${\om}$---see the related discussion in \cite{KrtousPodolsky:2004b}.}
along ${\scri}$.

In view of \eqref{lfixed}, with respect to the reference frame, a null direction
along a future oriented vector ${\kG}$ can be
parametrized by a spatial vector ${\Rv=R^i\,\mO{i}}$ which is orthogonal to ${\kO,\,\lO}$:
\begin{equation}\label{Rmeaning}
  \kG \propto \kO + \sqrt2\, \Rv + \abs{R}^2\, \lO\period
\end{equation}

Consequently, the null direction ${\kG}$ projected onto a space orthogonal to ${\tO}$ (cf.~(R5.5)) can be
represented by a unit spatial vector ${\qG}$
\begin{equation}\label{qproj}
  \qG = \frac1{1+\abs{R}^2}\bigl((1-\abs{R}^2)\,\qO + 2\,\Rv\bigr)\period
\end{equation}
The vector ${\Rv}$ is a thus a \emph{stereographic representation} of
the vector ${\qG}$, and hence of ${\kG}$. Indeed, if we introduce an angle ${\THT}$
between ${\qO}$ and ${\qG}$, and a unit direction ${\Ev}$ of the vector ${\Rv}$,
we obtain
\begin{equation}\label{RTHT}
  \qG=\cos\THT\;\qO+\sin\THT\;\Ev \comma\quad \Ev=\Rv/\abs{R}\comma\quad \abs{R}=\tan\frac\THT2\period
\end{equation}

Complementarily, a normalized projection ${\tG}$ of the null direction ${\kG}$
onto a timelike hypersurface ${\mathcal{H}_\refF}$
orthogonal to ${\qO}$ (cf.~(R5.8)) is given by
\begin{equation}\label{tproj}
  \tG = \frac1{\abs{1-\abs{R}^2}}\bigl((1+\abs{R}^2)\,\tO + 2\,\Rv\bigr)\period
\end{equation}
In this case, the vector ${\Rv}$ is a \emph{pseudostereographic representation} of the vector ${\tG}$.
In contrast to ${\qG}$, the vector ${\tG}$ does not represent a null
direction ${\kG}$ uniquely---the null direction obtained by reflection of ${\kG}$ with respect to
the hypersurface ${\mathcal{H}_\refF}$ leads to the same vector ${\tG}$. Therefore, we introduce
the sign ${\;\varsigma=\sign(1-\abs{R}^2)}$ which indicates if the vectors ${\kG}$ and ${\kO}$ have the
same orientation with respect to ${\mathcal{H}_\refF}$.
Introducing a rapidity parameter ${\PSI}$ between ${\tO}$ and ${\tG}$ we can write
\begin{equation}\label{RPSI}
  \tG=\cosh\PSI\;\tO+\sinh\PSI\;\Ev \comma\quad \Ev=\Rv/\abs{R}\comma\quad \abs{R}=\Bigl(\tanh\frac\PSI2\Bigr)^\varsigma\period
\end{equation}

Clearly, parametrization of the null direction ${\kG}$ using the vector ${\qG}$ and angle ${\THT}$
is useful for spacelike infinity ${\scri}$ where ${\norm\propto\tO}$ while the parametrization
using ${\tG}$, ${\PSI}$, and ${\varsigma}$ is more appropriate for timelike ${\scri}$ where ${\norm\propto\qO}$.

Finally, let us note that the null direction ${\kG\ant}$ \defterm{antipodal} to ${\kG}$, which is
defined by ${\qG\ant=-\qG}$, is given by ${\Rv\ant=-\Rv/\abs{R}^2}$,
and that the \defterm{mirrored} direction ${\kG\mir}$ obtained from ${\kG}$
by reflection with respect to the hypersurface ${\mathcal{H}_\refF}$ is
given by ${\tG\mir=\tG}$, ${\varsigma\mir=-\varsigma}$, so that ${\Rv=\Rv/\abs{R}^2}$.

\subsection{The interpretation frame}
\label{ssc:IntFrame}

By \defterm{interpretation frame} ${\kI,\,\lI,\,\mI{i}}$ we understand a null frame
that is parallelly transported along a null geodesic ${\geod(\afp)}$ to infinity ${\scri}$
with the vector ${\kI}$ tangent to the geodesic.
As in \cite{KrtousPodolsky:2004b}, we fix ${\kI=\frac1{\sqrt2}\frac{D\geod}{d\afp}}$.
Because we have already normalized the affine parameter ${\afp}$ by \eqref{InOutGeod},
this choice guarantees that both the geodesics and the interpretation frames approach
infinity from different directions in a comparable way.

The interpretation frame, however, is not uniquely fixed. One may perform
transformations which leave ${\kG}$ unchanged, namely the null rotation \eqref{kfixed}
and the spatial rotation \eqref{rotation}. This non-uniqueness
corresponds to the freedom in a choice of initial conditions for the frame.

A crucial observation which was first realized by Penrose is that
the interpretation frame (boosted by the conformal factor ${\om}$) \emph{becomes
adjusted} to infinity ${\scri}$, independently of its initial conditions.

This fact can be derived by comparing the boosted frame ${\kB=\om\kI}$, ${\lB=\om^{-1}\lI}$, ${\mB{i}=\mI{i}}$
with a frame parallelly transported in the conformal geometry.
Namely, we may define the auxiliary frame ${\kA,\,\lA,\,\mA{i}}$ as
a frame parallelly transported along the geodesic in the conformal geometry, isotropically rescaled
by ${\om}$ to become normalized in the physical geometry (see \cite{KrtousPodolsky:2004b} for a detailed discussion).
In addition, we require that the auxiliary frame is adjusted to infinity.
Following the steps leading to (R3.22), we analogously obtain that these two frames are related by
\begin{equation}\label{BAframerel}
  \kB = \kA\commae\quad
  \lB = \lA + \sqrt2 L^i \mA{i} + \abs{L}^2\, \kA\commae\quad
  \mB{i} = \Phi_i{}^j\,\bigl(\mA{j} + \sqrt2\,L_j\,\kA\bigr)\commae
\end{equation}
with parameters ${L^i(\afp)}$ and ${\Phi_i{}^j(\afp)}$ for
large affine parameter ${\afp}$ given by
\begin{equation}\label{PhiLinAfp}
   \Phi_i{}^j = \Phi_{(0)}{}_i{}^j\comma\quad
   L^i = -\EPS M_{(1)}^i\,\afp^{-1}\ln|\afp|+\EPS L_{(0)}^i\,\afp^{-1}+\ldots\period
\end{equation}
Here, ${L_{(0)}^i}$ and ${\Phi_{(0)i}{}^{j}}$ are constants of integration exactly corresponding
to the freedom in the choice of initial conditions for the interpretation frame.
The coefficients ${M_{(1)}^i}$ follow from the expansion of derivatives of ${\om}$
in the affine parameter ${\afp}$ along the geodesic, see equation~(R3.20). Typically---in the vacuum case
or for matter satisfying the asymptotic Einstein condition (R2.20)---${M_{(1)}^i}$
vanish, cf.\ eqation~(R3.23) and discussion therein.

In any case, we observe that
the boosted interpretation frame ${\kB,\,\lB,\,\mB{i}}$ at infinity
becomes equal to the auxiliary frame, i.e., it becomes adjusted to ${\scri}$, independently
of the parameter ${L_{(0)}^i}$. However, the dependence on the spatial
rotation ${\Phi_{(0)i}{}^{j}}$ persists.
Because (on the general level) we are not able to fix
the initial conditions for the interpretation frame uniquely,
we do not know a particular value of ${\Phi_{(0)i}{}^{j}}$
and its dependence on the direction of the geodesic. Therefore, we will
extract only information about radiation which is independent of the
choice of spatial rotation. For this reason, in the following we may ignore
any additional dependence on the spatial rotation.

We wish to characterize the interpretation frame with respect to the reference frame.
Both the auxiliary and the reference frame are adjusted to infinity, i.e. they
are related by a transformation which leaves the normal \eqref{Adjusted} unchanged.
If the direction ${\kA\propto\kI}$ is specified by the parameters ${R^i}$ via \eqref{Rmeaning},
the transformation from the reference to the auxiliary frame
can be obtained by consecutive application of the null rotation \eqref{kfixed} with ${\kG}$ fixed,
the null rotation \eqref{lfixed} with ${\lG}$ fixed, the boost \eqref{boost}, and
the spatial rotation \eqref{rotation}, given by parameters
\begin{equation}\label{ParamRefAux}
  L^i = \nsgn R^i\comma
  K^i = \frac{R^i}{1-\nsgn \abs{R}^2}\comma
  B = \EPS\EPS_\refF({1-\nsgn \abs{R}^2})\comma
\end{equation}
and some orthonormal matrix ${\Phi_i{}^j}$, which can be ignored.
The signs ${\EPS_\refF}$ and ${\EPS}$ indicate orientations of the vectors ${\kO}$ and ${\kA\propto\kI}$
with respect to infinity ${\scri}$, cf.\ eq.~\eqref{InOutGeod}.
Finally, the interpretation frame is simply obtained from the auxiliary one
by the boost \eqref{boost} with ${B=\om}$.

\section{The graviational field and its asymptotic structure}
\label{sc:field}

We want to analyze the asymptotic behaviour of the gravitational field.
For this we need to understand its algebraic structure.
However, in a dimension higher than 4, it is quite complicated.
It was investigated only recently in \cite{MilsonColeyPravdaPravdova:2005,ColeyMilsonPravdaPravdova:2004}.
Fortunately, these studies analyzed the structure of the Weyl tensor in the way which can
immediately be used to generalize our previous results \cite{KrtousPodolsky:2004b}.
We thus start with a short overview of the algebraic structure of the Weyl tensor
in higher dimensions. We will introduce a notation for its components which is convenient
for our purposes.

\subsection{Weyl tensor components and their transformation properties}
\label{ssc:fieldtrans}

We denote the frame components of the Weyl tensor as ${\WTc^{\refF}_{abcd}}$,
with index `${{}^\refF}$' indicating that the reference frame is considered. Inspired by the notation of
\cite{MilsonColeyPravdaPravdova:2005} we define the function ${\WTc^\refF_{abcd}\transf{l-fixed}{K^i}}$
of the argument ${K^i}$ as a Weyl tensor component ${\WTc_{abcd}}$ evaluated with respect of the frame
which is obtained from the reference frame by the null rotation \eqref{lfixed} with the parameters ${K^i}$.
Thanks to \eqref{lfixed} it is a polynomial in ${K^i}$ of the fourth order.
Similarly, we define the polynomials ${\WTc^\refF_{abcd}\transf{k-fixed}{L^i}}$,
${\WTc^\refF_{abcd}\transf{boost}{B}}$, and ${\WTc^\refF_{abcd}\transf{rotation}{\Phi_i{}^j}}$.
The explicit form of these functions can be easily obtained using equations~\eqref{kfixed}--\eqref{rotation},
but some of the expressions are rather cumbersome and we will not list them here.
Some particular transformations can be found in \eqref{WTcboost} and~\eqref{WTP4kfixed}.

The Weyl tensor frame components can be separated into various groups according to their transformation
properties under the boost \eqref{boost}. Any component ${\WTc_{abcd}}$ changes under
the boost in a very simple way, namely,
\begin{equation}  \label{WTcboost}
\WTc^\refF_{abcd}\transf{boost}{B} = B^{\weight(abcd)}\,\WTc^\refF_{abcd}\commae
\end{equation}
where the power ${\weight(abcd)}$ is called \defterm{boost weight} \cite{MilsonColeyPravdaPravdova:2005}.
For each set of frame indices ${a,b,c,d}$ it is simply
the number of indices corresponding to the vector ${\kO}$ minus
the number of indices corresponding to the vector ${\lO}$.
For the Weyl tensor, the boost weights take the values ${-2,-1,0,1,2}$. Components
with various boost weights ${w}$ are the analogues of the NP components ${\WTP{}{m}}$ of the Weyl tensor in
four dimensions with ${m=w+2}$. However, in a higher dimension, for each weight there are more than
two independent real components, so they cannot be
combined into suitable complex coefficients as when ${n=4}$.
Nevertheless, we may still introduce an analogous convenient notation,
and we distinguish different components by additional indices.\footnote{%
See the appendix for the relation to standard NP notation in the case ${n=4}$.}
We thus define real Weyl tensor components ${\WTP{}{m}_{\dots}}$ grouped by their boost weight ${w=m-2}$ as:
\begin{subequations}\label{WTPdef}
\begin{gather}
  \WTP{}{0}_{ij}\mspace{12.75mu}=\WTc_{abcd}\; k^a\, m_i^b\, k^c\, m_j^d\commae\\
  \WTP{}{1}_{ijk}\mspace{5mu}=\WTc_{abcd}\; k^a\, m_i^b\, m_j^c\, m_k^d\comma\mspace{26.5mu}
    \WTP{}{1\Tlbl}_{i}\mspace{7mu}=\WTc_{abcd}\; k^a\, l^b\, k^c\, m_i^d\commae\\
\begin{split}
  &\WTP{}{2}_{ijkl}=\WTc_{abcd}\; m_i^a\, m_j^b\, m_k^c\, m_l^d \comma\mspace{20mu}
    \WTP{}{2\Slbl}_{}\mspace{14.5mu}=2\WTc_{abcd}\; k^a\, l^b\, l^c\, k^d\commae\\
  &\WTP{}{2}_{ij}\mspace{12.5mu}=\WTc_{abcd}\;  k^a\, l^b\, m_i^c\, m_j^d\comma\mspace{38mu}
    \WTP{}{2\Tlbl}_{ij}=2\WTc_{abcd}\; k^a\, m_i^b\, l^c\, m_j^d\commae
\end{split}\\
  \WTP{}{3}_{ijk}\mspace{4.5mu}=\WTc_{abcd}\; l^a\, m_i^b\, m_j^c\, m_k^d\comma\mspace{31.5mu}
    \WTP{}{3\Tlbl}_{i}\mspace{6.75mu}=\WTc_{abcd}\; l^a\, k^b\, l^c\, m_i^d\commae\\
  \WTP{}{4}_{ij}\mspace{12.75mu}=\WTc_{abcd}\; l^a\, m_i^b\, l^c\, m_j^d\period
\end{gather}
\end{subequations}
%\hrule\vspace{1ex} {\sf old notation:}
%\addtocounter{equation}{-1}%
%\begin{subequations}
%\begin{gather}
%  \WTP{}{4}_{ij}\mspace{12.75mu}=\WTc_{abcd}\; l^a\, m_i^b\, l^c\, m_j^d\commae\\
%  \WTP{}{3}_{ijk}\mspace{4.5mu}=\WTc_{abcd}\; l^a\, m_i^b\, m_j^c\, m_k^d\comma\mspace{31.5mu}
%    \WTP{}{3}_{i}\mspace{18.5mu}=\WTc_{abcd}\; k^a\, l^b\, l^c\, m_i^d\commae\\
%\begin{split}
%  &\WTP{}{2}_{ijkl}=\WTc_{abcd}\; m_i^a\, m_j^b\, m_k^c\, m_l^d \comma\mspace{20mu}
%    \WTP{}{2}_{}\mspace{23.5mu}=\WTc_{abcd}\; k^a\, l^b\, k^c\, l^d\commae\\
%  &\WTP{}{2\Albl}_{ij}\mspace{1.25mu}=\WTc_{abcd}\;  k^a\, l^b\, m_i^c\, m_j^d\comma\mspace{38mu}
%    \WTP{}{2\Hlbl}_{ij}=\WTc_{abcd}\; k^a\, m_i^b\, l^c\, m_j^d\commae
%\end{split}\\
%  \WTP{}{1}_{ijk}\mspace{5mu}=\WTc_{abcd}\; k^a\, m_i^b\, m_j^c\, m_k^d\comma\mspace{26.5mu}
%    \WTP{}{1}_{i}\mspace{18.5mu}=\WTc_{abcd}\; l^a\, k^b\, k^c\, m_i^d\commae\\
%  \WTP{}{0}_{ij}\mspace{12.75mu}=\WTc_{abcd}\; k^a\, m_i^b\, k^c\, m_j^d\period
%\end{gather}
%\end{subequations}
%\hrule\vspace{1ex}
All other components can be obtained with the help of the symmetries of the Weyl tensor.
Moreover, the listed components are mutually related. The components
${\WTP{}{0}_{ij}}$, ${\WTP{}{1}_{ijk}}$, ${\WTP{}{2}_{ijkl}}$, ${\WTP{}{2}_{ij}}$,
${\WTP{}{3}_{ijk}}$, and ${\WTP{}{4}_{ij}}$ are independent up to constraints
following from the properties of the Weyl tensor:
\begin{subequations}
\begin{gather}
  \WTP{}{0}_{[ij]}=0\comma\WTP{}{0}_{k}{}^{k}=0\commae\\
  \WTP{}{1}_{i(kl)}=0\comma\WTP{}{1}_{[ikl]}=0\commae\\
  \WTP{}{2}_{ijkl}=\WTP{}{2}_{klij}\comma\WTP{}{2}_{(ij)kl}=\WTP{}{2}_{ij(kl)}=\WTP{}{2}_{i[jkl]}=0\comma\WTP{}{2}_{(ij)}=0\commae\\
  \WTP{}{3}_{i(kl)}=0\comma\WTP{}{3}_{[ikl]}=0\commae\\
  \WTP{}{4}_{[ij]}=0\comma\WTP{}{4}_{k}{}^{k}=0\period\label{psi4sym}
\end{gather}
\end{subequations}
The remaining components are not independent---they are given by
\begin{equation}
\begin{aligned}
  &\WTP{}{1\Tlbl}_{i}=\WTP{}{1}_{k}{}^{k}{}_{i}\commae\\
  &\WTP{}{2\Tlbl}_{ij}=\WTP{}{2}_{ikj}{}^{k}+\WTP{}{2}_{ij}\\
  &\WTP{}{2\Slbl}=\WTP{}{2\Tlbl}_{k}{}^{k}=\WTP{}{2}_{kl}{}^{kl}\commae\\
  &\WTP{}{3\Tlbl}_{i}=\WTP{}{3}_{k}{}^{k}{}_{i}\period
\end{aligned}
\end{equation}

Below we will mainly need the following transformation property of the ${\WTP{}{4}_{ij}}$
component under the null rotation \eqref{kfixed}:
\begin{equation}  \label{WTP4kfixed}
\begin{split}
  &\WTP{}{4}_{ij} = \WTP{\refF}{4}_{ij}\transf{k-fixed}{L^k}=\\
    &\;= \WTP{\refF}{4}_{ij} \\
    &\quad\;+ 2\sqrt2 \bigl( -\WTP{\refF}{3}_{(ij)k} L^k + \WTP{\refF}{3\Tlbl}_{(i} L_{j)} \bigr)\\
    &\quad\;+ \bigl( 2 \WTP{\refF}{2}_{ikjl} L^k L^l
                       - 2 \WTP{\refF}{2\Tlbl}_{k(i}L_{j)}L^k + \WTP{\refF}{2\Tlbl}_{(ij)} \abs{L}^2
                       - \WTP{\refF}{2\Slbl}_{} L_i L_j
                       - 4 \WTP{\refF}{2}_{k(i} L_{j)} L^k \bigr)\\
    &\quad\;-2\sqrt2 \bigl( 2 \WTP{\refF}{1}_{kl(i} L_{j)} L^k L^l + \WTP{\refF}{1}_{(ij)k} L^k \abs{L}^2
                            + \WTP{\refF}{1\Tlbl}_{(i} L_{j)} \abs{L}^2 - 2 \WTP{\refF}{1\Tlbl}_{k} L^k L_i L_j \bigr)\\
    &\quad\;+\bigl( 4 \WTP{\refF}{0}_{kl} L^k L^l L_i L_j -4 \WTP{\refF}{0}_{k(i}L_{j)} L^k \abs{L}^2 + \WTP{\refF}{0}_{ij} \abs{L}^4 \bigr)
    \period
\end{split}
\end{equation}
%\hrule\vspace{1ex} {\sf old notation:}
%\addtocounter{equation}{-1}%
%\begin{equation}
%\begin{split}
%  &\WTP{}{4}_{ij} = \WTP{\refF}{4}_{ij}\transf{k-fixed}{L^k}=\\
%    &\;= \WTP{\refF}{4}_{ij} \\
%    &\quad\;- 2\sqrt2 \bigl( \WTP{\refF}{3}_{(ij)k} L^k + \WTP{\refF}{3}_{(i} L_{j)} \bigr)\\
%    &\quad\;+ 2 \bigl( \WTP{\refF}{2}_{ikjl} L^k L^l
%                       - 2\WTP{\refF}{2\Hlbl}_{k(i}L_{j)}L^k + \WTP{\refF}{2\Hlbl}_{(ij)} \abs{L}^2
%                       + \WTP{\refF}{2}_{} L_i L_j
%                       - 2 \WTP{\refF}{2\Albl}_{k(i} L_{j)} L^k \bigr)\\
%    &\quad\;-2\sqrt2 \bigl( 2 \WTP{\refF}{1}_{kl(i} L_{j)} L^k L^l + \WTP{\refF}{1}_{(ij)k} L^k \abs{L}^2
%                            - \WTP{\refF}{1}_{(i} L_{j)} \abs{L}^2 + 2 \WTP{\refF}{1}_{k} L^k J_i L_j \bigr)\\
%    &\quad\;+\bigl( \WTP{\refF}{0}_{ij} \abs{L}^4 -4 \WTP{\refF}{0}_{k(i}L_{j)} L^k \abs{L}^2 + 4 \WTP{\refF}{0}_{kl} L^k L^l L_i L_j \bigr)
%    \period
%\end{split}
%\end{equation}
%\hrule\vspace{1ex}

\subsection{Weyl aligned null directions and algebraic classification in higher dimensions}
\label{ssc:WAND}

Following \cite{MilsonColeyPravdaPravdova:2005,ColeyMilsonPravdaPravdova:2004}, we call the null direction
${\kG}$ for which all the components ${\WTP{}{0}_{ij}}$ with respect
to the null frame ${\kG,\,\lG,\,\mG{i}}$ vanish \defterm{Weyl aligned null direction} (WAND).
This definition is independent of the choice of the normalization of ${\kG}$
and of the choice of other vectors ${\lG,\,\mG{i}}$ of the frame
because under transformations which leave
the direction ${\kG}$ unchanged, ${\WTP{}{0}_{ij}}$ behave as
\begin{equation}
\WTP{}{0}_{ij}\!\transf{k-fixed}{L^k}=\WTP{}{0}_{ij}\comma\!\!
\WTP{}{0}_{ij}\!\transf{boost}{B}=B^{2}\WTP{}{0}_{ij}\comma\!\!
\WTP{}{0}_{ij}\!\transf{rotation}{\Phi_k{}^l}=\Phi_i{}^k\,\Phi_j{}^l\;\WTP{}{0}_{kl}\period
\end{equation}
If, in addition, all the components ${\WTP{}{m}_{\dots}}$, ${m=0,\dots,o-1}$,
with respect the null frame ${\kG,\,\lG,\,\mG{i}}$ also vanish, WAND ${\kG}$
is said to be of \defterm{alignment order~${o}$}.
Again, such an order of the alignment depends only on the direction of ${\kG}$.

If we parametrize the null vector ${\kG}$ with respect to the reference frame
using the parameters ${R^k}$ according to \eqref{Rmeaning},
the conditions that ${\kG}$ is a WAND of alignment order~${o}$ become
\begin{equation}  \label{WANDdef}
   \WTP{\refF}{m}_{ij}\transf{l-fixed}{R^k}=0\qquad\text{for}\quad m=0,\dots,o-1\commae
\end{equation}
which are called \defterm{aligment equations} \cite{MilsonColeyPravdaPravdova:2005}.

In four-dimensional spacetimes WANDs always exist and are exactly the
principal null directions of the standard algebraic classification. In higher dimensions,
the alignment equation \eqref{WANDdef} even of the first alignment order can be too restrictive
and in a generic situation no WAND exists. If the alignment equations admit
solutions, the spacetime is called algebraically special. Such spacetimes
can be naturally classified according to the maximum alignment order of WANDs. The highest
alignment order ${o}$ of the WAND ${\kG}$ is called the \defterm{principal alignment type}.
For ${o=0,1,2,3,4}$, and ${5}$ we say that the spacetime is of principal type
G (general), I, II, III, N (null), and O (trivial), respectively.

Additionally, if we set the vector ${\kG}$ of the null frame ${\kG,\,\lG,\,\mG{i}}$
to be a WAND of the maximal alignment order, the highest possible aligment order of the vector ${\lG}$
is said to be of \defterm{secondary aligment type}.
We call the null frame with such chosen vectors ${\kG}$, ${\lG}$
the \defterm{frame aligned with the algebraic structure of the Weyl tensor}.
From the duality between the vectors ${\kG}$ and ${\lG}$
and the components ${\WTP{}{m}}$ and ${\WTP{}{4\!-\!m}}$, we conclude
that for a spacetime of principal and secondary alignment type ${p}$ and ${s}$
the Weyl tensor components ${\WTP{}{m}}$ in the aligned null frame
vanish for ${m=0,\dots,p-1}$ and ${m=4\!-\!s\!+\!1,\dots,4}$,
and the components ${\WTP{}{p}}$ and ${\WTP{}{4\!-\!s}}$ are nonvanishing.

Obviously, such a classification is a generalization of
the standard algebraic classification in four dimensions. The main difference is that
WANDs may not exist (i.e., the principal and secondary types can be zero).
In other words, in four dimensions there are allowed only types that are labeled by
principal and secondary type ${(p,s)}$ as ${(1,1)}$---Petrov type~I,
${(2,1)}$---Petrov type~II, ${(2,2)}$---Petrov type~D, ${(3,1)}$---Petrov type~III,
${(4,0)}$---Petrov type~N, and the trivial ${(5,5)}$---Petrov type~O.
In higher dimensions there are additional types ${(0,0)}$---type~G,
${(1,0)}$, ${(2,0)}$, and ${(3,0)}$; see \cite{MilsonColeyPravdaPravdova:2005,ColeyMilsonPravdaPravdova:2004}
for more detailed discussion.

\subsection{Asymptotic behaviour of the field components}
\label{ssc:fieldasympt}

Now we should specify behaviour
of the gravitational field at conformal infinity. However, it is not
our goal here to study a specific \vague{fall-off} of the field
in a general number of dimensions---such a task goes far beyond
the scope of this work. We are interested in the \emph{directional structure}
of the \vague{far} field and it turns out that qualitatively this structure is \emph{not}
affected by a specific fall-off property of the field.
We will thus make a general assumption that the Weyl tensor ${\WTc_{abcd}}$
behaves near infinity as ${\om^{q-2}}$, ${q}$ being some constant depending
on the dimension (and maybe even on a particular solution),\footnote{%
Alternatively, we could say that the fall-off of the
conformally related Weyl tensor is ${\cWT_{abc}{}^{d}\sim\om^q}$.}
cf.\ \cite{PravdovaPravdaColey:2005}.
In ${n=4}$ the well-known behaviour of the Weyl tensor is characterized by ${q=1}$.
For higher number of dimensions it is not clear if a similar property holds
in a general situation (see, e.g., \cite{HollandsWald:2004,HollandsIshibashi:2005}).
However, our assumption is rather \vague{weak} and we expect
it to be valid for a wide class of solutions.

Assuming the above fall-off of the Weyl tensor we can write down asymptotic
behaviour of its components. Combining \eqref{WTPdef}, \eqref{OmegaInAfp}
and considering the normalization of the reference frame\footnote{%
At ${\scri}$, the conformally rescaled frame ${\om^{-1}\kO,\om^{-1}\lO,\om^{-1}\mO{i}}$,
normalized in the unphysical metric ${\cmtrc}$, is regular.}
as ${\kO,\lO,\mO{i}\sim\om}$ we get
\begin{equation}\label{refasymp}
\WTP{\refF}{m}_{\dots}\lteq \WTPhat{\refF}{m}_{\dots}\,\afp^{-q-2}\commae
\end{equation}
where ${\WTPhat{\refF}{m}_{\dots}}$ are constant finite coefficients. We may also define
${\WTPhat{\refF}{m}_{\dots}\transf{k-fixed}{L^k}}$ in a similar way as
${\WTP{\refF}{m}_{\dots}\transf{k-fixed}{L^k}}$. It is a polynomial in
${L^k}$ with ${\WTPhat{\refF}{m}_{\dots}}$ being coefficients.

The relation between the interpreation and reference
frames was described in section \ref{ssc:IntFrame}.
In particular, the interpretation tetrad can be obtained by the sequence
of null rotations, boost, and spatial rotation with parameters
\eqref{ParamRefAux} followed by the (asymptotically singular) boost ${B=\om}$.
The field components with respect to the interpreation frame are thus
\begin{equation}  \label{radintaux}
\WTP{\intF}{m}_{\dots} \;\lteqrot\; \om^{2-m}\, \WTP{\auxF}{m}_{\dots} \comma
\end{equation}
since the interpretation frame is related to the adjusted auxiliary one by the boost.
Here and in the following, we ignore spatial rotations acting on indices ${i,j}$
which can arise from the non-uniqueness of the interpretation frame or from
the relation to the reference frame.
This will be indicated by a star `$*$' above the equality symbol.
Because asymptotically ${\om\sim\afp^{-1}}$, cf.\ \eqref{OmegaInAfp},
and the transformation from the
auxiliary frame to the reference frame is regular, we see that different
components of the Weyl tensor peel-off with different powers of the
affine parameter ${\afp}$ according to their boost weights labeled by ${m}$.
This is the analogue of the well-known peeling-off theorem in four dimensions
\cite{PenroseRindler:book} --- see also the recent analysis of this topic \cite{PravdovaPravdaColey:2005}
in higher dimensions.

We will now study only the leading component of the gravitational field which is ${\WTP{\intF}{4}_{ij}}$.
Performing the transformation \eqref{ParamRefAux} from the auxiliary frame to the reference frame
we obtain
\begin{equation}  \label{radauxref}
\WTP{\auxF}{4}_{ij} \;\lteqrot\; \frac{1}{(1-\nsgn \abs{R}^2)^2}\; \WTP{\refF}{4}_{ij}\transf{k-fixed}{\nsgn R^k} \commae
\end{equation}
where we parametrized the direction ${\kI}$ of the geodesic by ${R^k}$ via \eqref{Rmeaning}.
Using \eqref{radintaux}, \eqref{OmegaInAfp}, and \eqref{refasymp} we finally get
\begin{equation}  \label{radintref}
\WTP{\intF}{4}_{ij} \;\lteqrot\; \frac{\afp^{-q}}{(1-\nsgn \abs{R}^2)^2}\; \WTPhat{\refF}{4}_{ij}\transf{k-fixed}{\nsgn R^k} \period
\end{equation}
The explicit form of the expression ${\WTPhat{\refF}{4}_{ij}\transf{k-fixed}{L^k}}$ is given in \eqref{WTP4kfixed}.

\subsection{Directional structure of radiation}
\label{ssc:dirstrrad}

We have thus derived the asymptotic directional structure of gravitational
radiation in higher dimensions which is a generalization of the
main result of \cite{KrtousPodolsky:2004b}. It describes the dependence of the leading field component
on a direction along which the infinity is approached. Let us now briefly discuss
some properties of this dependence.

First, as we mentioned earlier, we have ignored an arbitrary rotation
of the field \eqref{radintref} \vague{in indices ${ij}$}. On the general level,
we cannot control the spatial rotation \eqref{rotation}
in these indices and therefore the only physically relevant quantities are
invariants which we can construct from the matrix \eqref{radintref}.
Thanks to the properties of the Weyl tensor, the matrix of the radiative field component ${\WTP{\intF}{4}_{ij}}$
is symmetric, cf.~\eqref{psi4sym}. Therefore, the invariants under spatial rotation
are real eigenvalues of the matrix. Alternative invariants are the traces of powers of the matrix.
Because ${\WTP{\intF}{4}_{ij}}$ is traceless, the independent invariants are
\begin{equation}  \label{invariants}
 \underset{\text{$m$-times}}{\underbrace{\;\WTP{\intF}{4}_{i}{}^{j}\;\; \WTP{\intF}{4}_{j}{}^{k}\;\dots\; \WTP{\intF}{4}_{l}{}^{i}}}
 \qquad\text{for}\quad m=2,\dots,n-2\period
\end{equation}
Substituting the relation \eqref{radintref} into \eqref{invariants}, one obtains expressions
which are polynomial in the directional parameter ${R^k}$.
Clearly, if all the invariants are zero, the complete leading term also vanishes.

The location of the zeros of the directional pattern \eqref{radintref} follows from a simple
argument. The pattern is proportional to ${\WTP{\auxF}{4}_{ij}}$.
Analogous to the alignment equations \eqref{WANDdef}, the vanishing of these Weyl tensor
components means that the null vector ${\lA}$ is asymptotically\footnote{%
The word \vague{asymptotically}
refers to the fact that we define WANDs at conformal infinity using the components ${\WTPhat{\refF}{m}_{\dots}}$
instead of ${\WTP{\refF}{m}_{\dots}}$,
i.e., using the Weyl tensor isotropically rescaled by a factor ${\sim \om^{-3}}$.
Such a definition of asymptotic WANDs is justified by the observation
that a finite isotropic rescaling of the Weyl tensor does not change the notion of WANDs.}
WAND. However, the auxiliary
frame is adjusted to infinity, i.e., the vectors ${\kA}$ and ${\lA}$ are related by
\eqref{Adjusted}. Hence, the direction of the geodesic ${\kI\propto\kA}$
along which the leading term of the field vanishes
has to be \vague{opposite} to a WAND ${\lA}$ in the sense of the relation \eqref{Adjusted}.
If the direction ${\kI\propto\kA}$ is given by the directional parameter ${R^k}$
through \eqref{Rmeaning}, such a direction ${\lA}$ is given by the parameter ${\nsgn^{-1} R^k/\abs{R}^2}$.

For ${\nsgn\neq0}$, it is also possible to reach the same conclusions using the identity
\begin{equation}  \label{Psi40}
  \WTPhat{\refF}{4}_{ij}\transf{k-fixed}{L^p} =
  \abs{L}^4\;\bigl(\krdel^k_i-2L_i L^k/\abs{L}^2\bigr)\bigl(\krdel^l_j-2L_j L^l/\abs{L}^2\bigr)\;
  \WTPhat{\refF}{0}_{kl}\transf{l-fixed}{L^p/\abs{L}^2}
\end{equation}
which relates transformations of the components ${\WTP{\refF}{4}_{ij}}$ and ${\WTP{\refF}{0}_{ij}}$.
Applying this identity to the directional structure \eqref{radintref},
and using the fact that matrix ${(\krdel^k_i-2L_i L^k/\abs{L}^2)}$ is orthogonal
we obtain
\begin{equation}
\WTP{\intF}{4}_{ij}
  \lteqrot \frac{\afp^{-q}\,\abs{R}^4}{(1-\nsgn \abs{R}^2)^2}\; \WTPhat{\refF}{0}_{ij}\transf{l-fixed}{\nsgn^{-1} R^k/\abs{R}^2}
  \comma\quad\text{for}\quad\nsgn\neq0\period
\end{equation}
Clearly, this vanishes iff the direction given by the parameter ${\nsgn^{-1} R^k/\abs{R}^2}$ is a WAND.

For ${\nsgn=0}$, equation \eqref{radintref} reduces to
\begin{equation}\label{radnull}
\WTP{\intF}{4}_{ij} \lteqrot \afp^{-q}\, \WTPhat{\refF}{4}_{ij}\period
\end{equation}
Thus ${\WTPhat{\refF}{4}_{ij}}$ vanishes if the vector ${\lO}$ is asymptotically WAND. However, ${\lO}$
is described by an infinite value of directional parameter \eqref{Rmeaning}, so that
the leading term ${\WTP{\intF}{4}_{ij}}$ again vanishes again iff the direction ${\nsgn^{-1} R^k/\abs{R}^2=\infty}$ is a WAND.

Let us now discuss the directional structure of radiation separately for
a different character of the conformal infinity ${\scri}$. For \emph{null character} of the infinity,
${\nsgn=0}$, we find that the leading term \eqref{radnull} is independent
of the direction of the null geodesic along which the infinity is approached.
It vanishes if the null direction ${\lO}$ tangent to the infinity (cf.\ adjustment condition
\eqref{Adjusted} for ${\nsgn=0}$) is asymptotically a WAND. This fact may be used for
an invariant characterization of the presence of gravitational radiation in higher-dimensional spacetimes.

For a \emph{spacelike} conformal infinity, ${\nsgn=-1}$, it is natural to parametrize
the null direction ${\kG}$ of the geodesic using its normalized projection ${\qG}$
to ${\scri}$, cf.\ equation~\eqref{qproj}. It can be expressed in terms of
the angle ${\THT}$ and of the complementary directional vector ${\Ev=e^k\mO{k}}$, see \eqref{RTHT}.
The directional structure \eqref{radintref} then reads
\begin{equation}
\WTP{\intF}{4}_{ij} \;\lteqrot\;
   \afp^{-q}\, \cos^4{\textstyle\frac\THT2}\;
   \WTPhat{\refF}{4}_{ij}\transf{k-fixed}{-\tan\!{\textstyle\frac\THT2}\;\, e^k} \period
\end{equation}
The leading term of the field vanishes if the direction ${\kG\ant}$ antipodal to ${\kG}$
is a WAND. Let us recall that the antipodal direction has the opposite projection
to ${\scri}$, ${\qG\ant=-\qG}$ (see the end of section \ref{ssc:RefFrame}).

For a \emph{timelike} infinity ${\scri}$, ${\nsgn=+1}$, we parametrize the null direction ${\kG}$
of the geodesic by its normalized projection ${\tG}$ to ${\scri}$
(unit timelike future oriented vector, cf.\ \eqref{tproj}) and by
the parameter ${\EPS=\pm1}$ which describes to which side of infinity the vector ${\kG}$ points
(cf.\ equation~\eqref{DefinitionEpsilon}). If we express the vector ${\tG}$ through the rapidity ${\PSI}$
and the directional vector ${\Ev}$, see equation~\eqref{RPSI}, we obtain
\begin{equation}
\WTP{\intF}{4}_{ij} \;\lteqrot\;
   \afp^{-q}\, {\textstyle\frac14} (\cosh\PSI+\EPS\EPS_\refF)^2\;
   \WTPhat{\refF}{4}_{ij}\transf{k-fixed}{\tanh^{\EPS\EPS_\refF}\!\!{\textstyle\frac\PSI2}\;\, e^k} \period
\end{equation}
This vanishes if the mirror reflection ${\kG\mir}$ of the direction ${\kG}$ with respect to the infinity
is a WAND (see again the end of section \ref{ssc:RefFrame}).

We conclude that all these results are direct generalizations of the analogous results of \cite{KrtousPodolsky:2004b}.
As in four dimensions, the directional structure of radiation is given by the algebraic structure
of the Weyl tensor. Also, the directions of vanishing radiation are determined by WANDs which are
generalizations of the principal null directions which are known from ${n=4}$ general relativity.

\subsection{Algebraically special spacetimes}
\label{ssc:algspecST}

The general explicit form of the directional dependence of the radiative component is rather cumbersome,
cf.\ equation~\eqref{radintref} combined with \eqref{WTP4kfixed}.
It simplifies for algebraically special
spacetimes, i.e., in spacetimes which posses some WANDs, see section \ref{ssc:WAND}.
Let us emphasize, however, that in higher dimensions
the condition that the spacetime is algebraically special can be rather
restrictive---as we mentioned above, a generic spacetime has no WANDs.
It is also not clear if an algebraically special spacetime can admit a regular global infinity ${\scri}$.
Fortunately, our discussion requires only the \emph{local} existence and regularity of the conformal infinity,
and we restrict only to the cases when such an infinity exists.

Although the directional structure of radiation simplifies for algebraically special spacetimes,
it is still parameterized by more independent components of the Weyl tensor
than in the case of four dimensions. The resulting expressions thus typically remain lengthy.
We will present them only in two most special cases---for  spacetimes of type~N and of type~III.

A substantial simplification of the directional structure \eqref{radintref} occurs only
for maximally special (nontrivial) spacetimes of type~N. In this case there exists
a WAND of the alignment order~4. If we choose the reference tetrad in such a way that
${\kO}$ is asymptotically this WAND, the alignment equations tell us that only the components ${\WTPhat{\refF}{4}_{ij}}$
are nonvanishing. In view of \eqref{WTP4kfixed}, the directional structure of the radiative field components thus takes a simple form
\begin{equation}\label{radN}
\begin{split}
\WTP{\intF}{4}_{ij} \;&\lteqrot\;
  \frac{\afp^{-q}}{(1-\nsgn \abs{R}^2)^2}\; \WTPhat{\refF}{4}_{ij}\\
  \;&\lteqrot\; \afp^{-q}\, \cos^4{\textstyle\frac\THT2}\; \WTPhat{\refF}{4}_{ij}
  \;\lteqrot\; \afp^{-q}\, {\textstyle\frac14} (\cosh\PSI+\EPS\EPS_\refF)^2\; \WTPhat{\refF}{4}_{ij}\period
\end{split}
\end{equation}
Here ${\WTPhat{\refF}{4}_{ij}}$ are constants characterizing the \vague{strength} and \vague{polarization}
of the field.
In this case, it is more convenient to choose as invariants (which ignore unknown polarization) the eigenvalues of
${\WTP{\intF}{4}_{ij}}$ instead of the traces \eqref{invariants}---these
are proportional to the eigenvalues of ${\WTPhat{\refF}{4}_{ij}}$
with a common factor which can be read from equation~\eqref{radN}.

The next simplest case is that of spacetimes of type~III
(we do not need to distinguish the type
according the secondary alignment type in our discussion).
In this case, we may choose the reference tetrad with ${\kO}$ pointing asymptotically
along a WAND of alignment order~3. It follows that the
components ${\WTPhat{\refF}{m}_{\dots}}$, ${m=0,1,2}$, are vanishing
and the leading field component thus reads
\begin{equation}  \label{radIII}
\WTP{\intF}{4}_{ij} \;\lteqrot\; \frac{\afp^{-q}}{(1-\nsgn \abs{R}^2)^2}\;
    \Bigl(\WTPhat{\refF}{4}_{ij}
    + \nsgn\,2\sqrt2\, \bigl( - \WTPhat{\refF}{3}_{(ij)k} R^k + \WTPhat{\refF}{3\Tlbl}_{(i} R_{j)} \bigr)
    \Bigr)\period
\end{equation}
It would be possible to use a more detailed structure of the
field components ${\WTPhat{\refF}{4}_{ij}}$ and ${\WTPhat{\refF}{3}_{ijk}}$
to select a \vague{canonical} reference frame (e.g., ${\WTPhat{\refF}{3\Tlbl}_{i}}$ identifies
an additional spatial direction, etc.) with respect to which \eqref{radIII} would have
a slightly more specific form. However, such a discussion
would not bring any significant additional understanding of the directional
structure of the field and we will not enter it here.

\section{Conclusion}

In recent years a great effort has been devoted to the investigation of
gravitational theories in higher dimensional spacetimes.
Several exact solutions of (generalized) Einstein's equations
with properties either analogous to the four-dimensional case
or with completely new features
(such as, e.g., the existence of black rings) were found.
However, many useful concepts and methods known from the four-dimensional gravity
still have not been generalized to higher dimensions.

Our work is a contribution to such possible generalizations, namely to
a discussion of an asymptotic behaviour of the gravitational
field in higher dimensions. It does not address in detail such questions
as what exactly radiation is or which part of the gravitational field
is relevant for physical observers (e.g., in brane-world scenarios
we should restrict only to the part of the conformal infinity near the brane).
However, it demonstrates that the structure of the leading field
components---in the sense of the peeling theorem---can be described in an analogous way as in
four dimensions. It shows, that the directional ambiguity
of the leading components in the case of a non-vanishing cosmological constant
can again be characterized in terms of the asymptotic algebraic structure
of the Weyl tensor.
Due to the more complicated algebraic properties of the Weyl tensor
in higher dimensions, the directional structure of the radiative components
is, not surprisingly, more intricate.

\appendix

\section{Relation to complex notation in ${n=4}$}

In standard ${n=4}$ general relativity it is convenient to introduce a \emph{complex null tetrad}
and to parametrize the Weyl tensor by the corresponding five \emph{complex} components. These NP quantities are
closely related to the \emph{real} quantities introduced in our text. Here we
present a \vague{dictionary} relating these two notations.

In four dimensions the transverse indices ${i,j,k,l}$ run only over two values ${1,2}$ and
we can combine the real vectors ${\mG{i}}$ into the complex vectors
\begin{equation}
\cG={\textstyle\frac1{\sqrt2}}\,(\mG{1}-i\mG{2})\comma \bG={\textstyle\frac1{\sqrt2}}\,(\mG{1}+i\mG{2})\period
\end{equation}
Any real vector ${\tens{V}}$ spanned on ${\mG{1},\mG{2}}$ can be parametrized by a complex number ${V}$ by
the relation
\begin{equation}
\tens{V}=V^1\mG{1}+V^2\mG{2}={\textstyle\frac1{\sqrt2}}\,(\bar V\cG+V\bG)\period
\end{equation}
It follows that
\begin{equation}
V=V^1-iV^2\comma\abs{V}^2=(V^1)^2+(V^2)^2=V\bar{V}\period
\end{equation}
The transformation properties of the null tetrad under, for example, a null rotation with ${\kG}$ fixed \eqref{kfixed}
then reads
\begin{equation}
  \kG = \kO\commae\qquad
  \lG = \lO + \bar{L}\, \cG_{\refF} + L\, \bG_{\refF} +  \abs{L}^2\, \kO\commae\qquad
  \cG = \cG_{\refF} + L\,\kO\period
\end{equation}
This is the standard four-dimensional expression  for a null rotation, see, e.g., \cite{KrtousPodolsky:2004b}.

Moreover, in four dimensions there are only two real independent components of the Weyl tensor
for each boost weight, namely
\begin{subequations}
\begin{gather}
\WTP{}{0}_{11}=-\WTP{}{0}_{22}\comma\WTP{}{0}_{12}=\WTP{}{0}_{21}\commae\\
\WTP{}{1\Tlbl}_{1}=\WTP{}{1}_{221}=-\WTP{}{1}_{212}\comma
\WTP{}{1\Tlbl}_{2}=\WTP{}{1}_{112}=-\WTP{}{1}_{121}\commae\\
\begin{split}
&\WTP{}{2}_{1212}=\WTP{}{2}_{2121}=-\WTP{}{2}_{2112}=-\WTP{}{2}_{1221}=\WTP{}{2\Tlbl}_{11}=\WTP{}{2\Tlbl}_{22}={\textstyle\frac12}\WTP{}{2\Slbl}\commae\\
&\WTP{}{2}_{12}=-\WTP{}{2}_{21}=\WTP{}{2\Tlbl}_{12}=-\WTP{}{2\Tlbl}_{21}\commae
\end{split}\\
\WTP{}{3\Tlbl}_{1}=\WTP{}{3}_{221}=-\WTP{}{3}_{212}\comma
\WTP{}{3\Tlbl}_{2}=\WTP{}{3}_{112}=-\WTP{}{3}_{121}\commae\\
\WTP{}{4}_{11}=-\WTP{}{4}_{22}\comma\WTP{}{4}_{12}=\WTP{}{4}_{21}\period
\end{gather}
\end{subequations}
These can be combined into complex NP components defined by
\begin{subequations}
\begin{align}
  \WTP{}{0} &=  \WTc_{abcd}\, k^a \,m^b \,k^c \,m^d \comma\\
  \WTP{}{1} &=  \WTc_{abcd}\, k^a \,l^b \,k^c \,m^d \commae\\
  \WTP{}{2} &=  \WTc_{abcd}\, k^a \,m^b \,\bar{m}^c \,l^d \commae\\
  \WTP{}{3} &=  \WTc_{abcd}\, l^a \,k^b \,l^c \,\bar{m}^d \comma\\
  \WTP{}{4} &=  \WTc_{abcd}\, l^a \,\bar{m}^b \,l^c \,\bar{m}^d \commae
\end{align}
\end{subequations}
via
\begin{subequations}
\begin{align}
  \WTP{}{0} &=  \WTP{}{0}_{11}-i\WTP{}{0}_{12} \comma\\
  \WTP{}{1} &=  {\textstyle\frac1{\sqrt2}} (\WTP{}{1\Tlbl}_{1}-i\WTP{}{1\Tlbl}_{2}) \commae\\
  \WTP{}{2} &=  -(\WTP{}{2}_{1212}+i\WTP{}{2}_{12}) \comma\\
  \WTP{}{3} &=  {\textstyle\frac1{\sqrt2}} (\WTP{}{3\Tlbl}_{1}+i\WTP{}{3\Tlbl}_{2}) \commae\\
  \WTP{}{4} &=  \WTP{}{4}_{11}+i\WTP{}{4}_{12} \period
\end{align}
\end{subequations}

%\bibliographystyle{myiop}
%\bibliography{D:/odborne/library/TeX/bib/references}
%\end{document}

\section*{References}

\enlargethispage{5ex}

\end{document}